\documentclass[aps,prc,twocolumn,groupedaddress,nofootinbib]{revtex4-2}

\usepackage{amsmath,amssymb,multirow,graphicx,color}
\def\Vec#1{\boldsymbol{#1}}

\begin{document}


\title{
Core structures of vortices in
Ginzburg-Landau theory for 
neutron $^3P_2$ superfluids
}


\author{Michikazu Kobayashi}
\affiliation{School of Environmental Science and Engineering, Kochi University of Technology, Miyanoguchi 185, Tosayamada, Kami, Kochi 782-8502, Japan}

\author{Muneto Nitta}
\affiliation{Department of Physics, and Research and Education Center for Natural Sciences, Keio University, Hiyoshi 4-1-1, Yokohama, Kanagawa 223-8521, Japan}


\date{\today}

\begin{abstract}
We investigate vortex solutions 
in the Ginzburg-Landau theory 
for neutron $^3P_2$ superfluids relevant for neutron star cores  
 in which neutron pairs possess the total angular momentum $J=2$ with spin-triplet and $P$ wave, 
in the presence of 
the magnetic field parallel to the angular momentum of 
vortices. 
The ground state is known to be in the uniaxial nematic (UN) phase
in the absence of magnetic field, 
while it is in the $D_2$ ($D_4$) biaxial nematic (BN) 
phase in the presence of the magnetic field 
below (above) the critical value. 
We find that a singly quantized vortex always splits into two half-quantized non-Abelian vortices connected by soliton(s) 
as a vortex molecule  
with any strength of the magnetic field.
In the UN phase, 
two half-quantized vortices with 
ferromagnetic cores 
are connected by a linear soliton with 
the $D_4$ BN order.
In the $D_2$ ($D_4$) BN phase, 
two half-quantized vortices with cyclic cores are connected by
three linear solitons with the $D_4$ ($D_2$) BN order. 
The energy of the vortex molecule monotonically increases 
and 
the distance between the two half-quantized vortices 
decreases with the magnetic field increases,  
except for a discontinuously increasing  jump of the distance 
at the critical magnetic field. 
We also construct an isolated half-quantized non-Abelian vortex 
in the $D_4$ BN phase.

\end{abstract}


\maketitle

\section{Introduction}

Pulsars or rapidly rotating neutron stars are dense and compact stars under extreme conditions, thereby serving as astrophysical laboratories for studying nuclear and QCD matter 
at high density,  with a strong magnetic field and 
under rapid rotation 
\cite{Graber:2016imq,Baym:2017whm}.  
The recent progresses in astrophysical observations promote us to study the interiors of neutron stars more precisely: 
the observation of massive neutron stars whose masses are about twice as large as the solar mass~\cite{Demorest:2010bx,Antoniadis1233232}, 
the gravitational waves from a binary neutron star merger~\cite{TheLIGOScientific:2017qsa,Abbott:2020uma}, 
and 
the Neutron Star Interior Composition Explorer (NICER) mission~\cite{Riley:2019yda,
  Miller:2019cac},  
expected to reveal interior states of neutron stars. 

The interior of neutron stars is believed to exhibit neutron superfluidity and proton superconductivity 
as first predicted by Migdal \cite{Migdal:1960} 
(see 
Refs.~\cite{Chamel:2008ca,Chamel2017,Haskell:2017lkl,Sedrakian:2018ydt,Graber:2016imq,Andersson2021} 
for recent reviews). 
Such superfluid and superconducting components can alter 
low-energy excitation modes compared with the normal phase, 
and thus their existence can affect several processes and properties 
of neutron stars, such as neutrino emissivities and specific heats relevant to the long relaxation time after the sudden speed-up events, that is pulsar glitches, of neutron stars~\cite{Baym1969,Pines1972,Takatsuka:1988kx}, and the enhancement of neutrino emission around 
the critical point of the superfluid transition~\cite{Yakovlev:2000jp,Potekhin:2015qsa,Yakovlev:1999sk,Heinke:2010cr,Shternin2011,Page:2010aw}. 
The neutron superfluids are realized by the attraction between two neutrons  in the $^1S_0$ channel 
at the low density \cite{Migdal:1960}. 
It was, however, shown in Ref.~\cite{1966ApJ...145..834W} that this channel is repulsive 
at higher densities 
as a consequence of the strong short-range repulsion. 
Thus, it was proposed that 
neutron $^3P_2$ superfluids, in which neutron pairs possess the total angular momentum $J=2$ with spin triplet and $P$ wave, are more relevant at higher density~\cite{Tabakin:1968zz,Hoffberg:1970vqj,Tamagaki1970,Hoffberg:1970vqj,Takatsuka1971,Takatsuka1972,Fujita1972,Richardson:1972xn,Amundsen:1984qc,Takatsuka:1992ga,Baldo:1992kzz,Elgaroy:1996hp,Khodel:1998hn,Baldo:1998ca,Khodel:2000qw,Zverev:2003ak,Maurizio:2014qsa,Bogner:2009bt,Srinivas:2016kir}. 
The $^3P_2$ interaction originates from a strong spin-orbit 
force at large scattering energy, and thus the neutron $^3P_2$ superfluids are expected to be realized in the high-density regions in the inner cores of neutron stars. 
They also can survive in neutron stars with strong magnetic fields, 
such as magnetars with the magnetic field $10^{15}$--$10^{18}$ G, 
since they are tolerant against the strong magnetic field  
due to the fact that the aligned pairs of Cooper pairs with the spin-triplet pairing are not broken by the Zeeman effect. 
In the $S$-wave case, it has also predicted that Cooper pairs can survive at around the magnetic field $10^{17}$ G \cite{Stein:2016}.

In astrophysical observations,
the possibility of the existence of neutron $^{3}P_{2}$ superfluids inside neutron stars are investigated;
the rapid cooling of the neutron star in Cassiopeia A 
might be explained by the enhancement of neutrino emissivities due to the formation and dissociation of neutron $^3P_2$ Cooper pairs~\cite{Heinke:2010cr,Shternin2011,Page:2010aw}. 

On the other hand, 
as the theoretical aspects are concerned, 
neutron $^3P_2$ superfluids have rich topological structures 
both in bosonic and fermionic excitations. 
There are basically two related approaches for theoretical study 
of superfluids.
The most fundamental theory is 
the Bogoliubov-de Gennes (BdG) equation 
offering a framework 
to deal with fermion degrees of freedom.  
The other is 
the Ginzburg-Landau (GL) approach 
as the low-energy effective theory
obtained by integrating out fermion degrees of freedom,
which is an expansion of both the order parameters and spatial derivatives, 
and thus is valid only in the region close to the critical temperature. 
The BdG approach was applied to $^3P_2$ superfluids 
and the phase diagram in the plane of 
the temperature and magnetic field 
was obtained \cite{Mizushima:2016fbn}. 
Furthermore, $^3P_2$ superfluids 
were shown to be topological superfluids 
of a class DIII in the classification of topological insulators and superconductors~\cite{Schnyder:2008tya,Ryu:2010zza}, 
allowing a topologically protected 
gapless Majorana fermion on its boundary 
\cite{Mizushima:2016fbn}.
On the other hand,
within the GL theory,   
superfluid states with $J=2$ are in general  classified into nematic, cyclic, and ferromagnetic phases etc~\cite{merminPRA74}.
The GL theory for $^3P_2$ superfluids was obtained~\cite{Fujita1972,Richardson:1972xn,
Sauls:1978lna,Muzikar:1980as,Sauls:1982ie,Vulovic:1984kc,Masuda:2015jka,Masuda:2016vak, 
Yasui:2018tcr,Yasui:2019tgc,Yasui:2019unp},
and 
in the weak coupling limit,  
the nematic phase was found to be the ground state of $^3P_2$ superfluids~\cite{
Sauls:1978lna,Muzikar:1980as,Sauls:1982ie}.
The nematic phase consists of three subphases 
with different unbroken symmetries: 
uniaxial nematic (UN), $D_2$ biaxial nematic ($D_2$ BN), 
and $D_4$ biaxial nematic ($D_4$ BN) phases,
with unbroken groups $O(2)$, $D_2$ and $D_4$, respectively,  
where $D_n$ is a dihedral group of order $n$ 
[see Table~\ref{table}(a) and \ref{table}(b)].
Corresponding order parameter manifolds (OPMs) are 
$U(1) \times SO(3)/O(2) \simeq S^1 \times {\mathbb R}P^2$, 
$U(1) \times SO(3)/D_2$, and $[U(1) \times SO(3)]/D_4$, 
respectively 
[see Table~\ref{table}(c)].  
These are continuously degenerated 
in the absence of magnetic field in the GL expansion up to the fourth order \footnote{ 
More precisely 
these are connected by a parameter of continuous degeneracy 
called a quasi-Nambu-Goldstone mode 
\cite{Uchino:2010pf}, 
and these OPMs are 
submanifolds of an extended OPM 
$(S^1 \times S^4)/{\mathbb Z}_2$.
}.
In the presence of the magnetic field and/or 
with the inclusion of the sixth-order term into the GL theory,
the continuous degeneracy is lifted 
to pick up 
either UN, $D_2$ BN, or $D_4$ BN state as 
the ground state 
for zero magnetic field, nonzero one below the critical value $B_c$, 
and nonzero one above $B_c$, respectively 
\cite{Masuda:2015jka,Yasui:2018tcr,Yasui:2019tgc} 
[see Table~\ref{table}(a)]. 
There is a subtle problem on the instability 
of the ground states for large value of the order parameter, 
which is cured by the expansion up to 
the eighth order \cite{Yasui:2019unp}. 
The phase diagram up to the eighth order 
captures the essential features 
of that determined in the BdG equation \cite{Mizushima:2016fbn}, 
including a tricritical point connecting 
first- and second-order phase transition lines 
 between $D_4$ and $D_2$ BN phases 
\cite{Mizushima:2016fbn,Mizushima:2019spl}.  
Apart from nematic phases,
more general uniform states of $^3P_2$ superfluids were classified 
\cite{Kobayashi:2021arv}, 
which is also useful to identify local states such as 
vortex cores. 
As a uniform ground state, the ferromagnetic state is in fact found to appear, 
beyond the quasi-classical approximation, 
in the region close to the critical temperature 
\cite{Mizushima:2021qrz}. 
The GL approach is useful to deal with 
bosonic collective excitations and topological defects.  
 Bosonic excitations 
  in the $^3P_2$ superfluids 
 yield collective modes 
\cite{Bedaque:2003wj,Leinson:2011wf,Leinson:2012pn,Leinson:2013si,Bedaque:2012bs,Bedaque:2013rya,Bedaque:2013fja,Bedaque:2014zta,Leinson:2009nu,Leinson:2010yf,Leinson:2010pk,Leinson:2010ru,Leinson:2011jr} 
relevant, for instance, for cooling process of neutron stars.  
Topological defects such as
domain walls \cite{Yasui:2019vci} 
and 
the boundary defect (boojums) of $^3P_2$ superfluids~\cite{Yasui:2019pgb} were investigated.

One of the most salient features of superfluidity 
is the fact that circulations of vortices 
are quantized so that the wave function is single valued 
around the vortices  
(the Feynman-Onsager's quantization), 
yielding the existence of quantized vortices. 
When a superfluid is rotating, 
a vortex lattice is formed as observed in helium superfluids 
and ultracold atomic gasses. 
In the context of superfluids in neutron stars, 
the origin of pulsar glitches was proposed to be explained 
by sudden releases of a large number of quantized vortices~\cite{reichley,Anderson:1975zze}.
In the case of $^3P_2$ superfluids, 
quantized vortices were investigated both in the GL theory 
\cite{Muzikar:1980as,Sauls:1982ie,Fujita1972,Masuda:2015jka,Chatterjee:2016gpm,Masuda:2016vak} 
(see also Ref.~\cite{Leinson:2020xjz} for coreless vortices), 
and in the BdG theory 
\cite{Masaki:2019rsz,Masaki:2021hmk}.
The first homotopy group classifying vortices is given in 
Ref.~\cite{Masuda:2015jka} [see Table \ref{table}(d)].
Singly quantized vortices in $^3P_2$ superfluids 
were studied in the GL theory \cite{Muzikar:1980as,Sauls:1982ie,Fujita1972,Masuda:2015jka,Chatterjee:2016gpm}, 
and topologically protected Majorana fermions in 
the vortex core 
were found in the BdG theory \cite{Masaki:2019rsz}.
Vortices more peculiar to 
the $^3P_2$ superfluids 
are half-quantized non-Abelian vortices 
which are allowed only in the $D_4$ BN phase 
\cite{Masuda:2016vak,Masaki:2021hmk}.
Their circulations are 
a half of the Feynman-Onsager's quantized circulations, 
and 
 the first homotopy group characterizing these vortices 
 is non-Abelian, thus giving non-commutativity of 
 exchanging vortices. 
 The existence of half-quantized vortices 
was proposed to explain a scaling law of 
pulsar glitches \cite{Marmorini:2020zfp}.
In Ref.~\cite{Masaki:2021hmk}, 
it was found in the BdG formalism that 
a singly quantized vortex is split into 
two half-quantized non-Abelian vortices. 
It was also found that a Majorna fermion zero mode is trapped in 
each half-quantized vortex.  

Apart from $^3P_2$ superfluids, 
spin-2 spinor ultracold atomic Bose-Einstein condensates (BECs) 
are also $J=2$ condensates 
whose ground states are possibly nematic phase  
\cite{Zhou:2006fd,Semenoff:2006vv,
Uchino:2010,Uchino:2010pf,Borgh:2016cco,Kobayashi:2021arv} 
sharing almost the same bosonic properties with $^3P_2$ superfluids, 
thus admitting the same order parameter manifold and 
non-Abelian half-quantized vortices
\cite{Uchino:2010pf,Borgh:2016cco}.
Therefore, studying bosonic properties of 
$^3P_2$ superfluids is also applicable to spin-2 BECs, 
which can be experimentally testable in principle,
although the current experiments of $^{87}$Rb atoms imply their 
ground state to be in the cyclic or nematic phase 
\cite{Schmaljohann:2004,Chang:2004,Kuwamoto:2004,Widera:2006,Tojo:2008,Tojo:2009}. 

\begin{table}
\begin{tabular}{c||c|c|c}
 $|\Vec{B}| $   &  $|\Vec{B}| =0$ &  $0< |\Vec{B}| <B_c$ & $B_c < |\Vec{B}| $\\ \hline 
\multicolumn{1}{l||}{(a) Phase} & UN & $D_2$ BN & $D_4$ BN \\
\multicolumn{1}{l||}{(b) Symmetry} & $O(2)$ & $D_2$ & $D_4$ \\
\multicolumn{1}{l||}{(c) OPM} &
 $S^1 \times {\mathbb R}P^2$ 
& $U(1) \times \frac{SO(3)}{D_2}$ 
& $\frac{U(1) \times SO(3)}{D_4}$\\
\multicolumn{1}{l||}{(d) $\pi_1$(OPM)} 
& ${\mathbb Z} \oplus {\mathbb Z}_2$ 
& ${\mathbb Z} \oplus {\mathbb Q}$ 
& ${\mathbb Z} \rtimes_h D_4^* $\\ 
\multicolumn{1}{l||}{(e) Vortex core order} & Ferro & Cyclic & Cyclic \\
\multicolumn{1}{l||}{(f) $\#$ of solitons} & 1 & 3 & 3\\
\multicolumn{1}{l||}{(g) Soliton core order} &  $D_4$ BN & $D_4$ BN & $D_2$ BN
\end{tabular}
\caption{\label{table}
In the absence and presence of the magnetic field 
below and above the critical value $B_c$, 
(a) phases, (b) unbroken symmetries, 
(c) OPMs, (d) the first homotopy groups 
of the OPMs,
(e) the order inside half-quantized vortex cores, 
(f) the number of solitons connecting 
the two  half-quantized vortices, and 
(g) the order inside soliton cores are summarized.
Here, 
${\mathbb Q} =  D_2^* $ is a quaternion group 
with $*$ implying the universal covering group.
 $\rtimes_h$ is a product defined in Ref.~\cite{Kobayashi:2011xb},
 ensuring the existence of isolated half-quantized non-Abelian vortices.
(e), (f), and (g) are new results presented in this paper.
}
\end{table}

In this paper, we investigate vortex solutions, namely 
singly quantized vortices 
and half-quantized non-Abelian vortices, in neutron $^3P_2$ superfluids within the GL approach.
The orientation of the magnetic field is fixed to be parallel to the angular momentum of vortices.
In the previous studies of vortices in the GL theory,
an axial symmetry around the vortex axis was assumed 
\cite{Muzikar:1980as,Sauls:1982ie,Fujita1972,Masuda:2015jka,Chatterjee:2016gpm,Masuda:2016vak}.
In contrast, 
imposing no axial symmetry, we find that a singly quantized vortex always splits into two half-quantized non-Abelian vortices 
with any strength of the magnetic field.
An advantage to use the GL theory compared with 
the BdG equation employed in Ref.~\cite{Masaki:2019rsz} is 
that we do not have to consider the direction 
of splitting \textit{a priori}.
In the UN phase with the zero magnetic field,
cores of two half-quantized vortices 
are found to be filled with the
ferromagnetic states,
while
they are filled with the
cyclic states  in the $D_2$ and $D_4$ BN phases 
in the presence of the magnetic field,  
as summarized in Table \ref{table}(e).
In the UN phase, 
the two half-quantized vortices are connected by 
a single soliton of the $D_4$ BN order while 
they are connected by three linear solitons 
with the $D_4$ ($D_2$) BN order 
in the $D_2$ ($D_4$) BN phase,
as summarized in Table \ref{table}(f) and \ref{table}(g).
The appearance of the $D_4$ BN order around the vortex core 
in the UN and $D_2$ BN phases 
is a consequence 
of the fact that isolated half-quantized vortices can topologically exist only in the $D_4$ BN state. 
We also show that 
the energy of the vortex molecule monotonically increases 
as the magnetic field increases, 
which is continuous everywhere including the critical magnetic field 
separating $D_2$ and $D_4$ BN states. 
On the other hand, 
the distance between the two half-quantized vortices 
decreases with the magnetic field increases,  
except for a discontinuously increasing  jump 
at the critical magnetic field. 
We also construct an isolated half-quantized non-Abelian vortex 
in the $D_4$ BN phase 
above the critical magnetic field.

A molecule of half-quantized vortices 	connected by a soliton 
or domain wall can be found in various systems 
such as 
multicomponent or multigap superconductors 
\cite{Babaev:2001hv,Babaev:2004rm,Goryo2007,Tanaka2007,Crisan2007,Guikema2008,Nitta:2010yf,
Garaud:2011zk,
Garaud:2012pn,
Garaud2012a,
Tanaka2017,Tanaka2018,Chatterjee:2019jez},
coherently coupled multicomponent BECs \cite{Son:2001td,
Mueller2002,Kasamatsu2003,
Kasamatsu:2004tvg,Kuopanportti2012,Aftalion2012,Eto:2012rc,Cipriani:2013nya,Eto:2013spa,
Nitta:2013eaa,Dantas2015,Tylutki:2016mgy,Eto:2017rfr,Eto:2019uhe,Kobayashi:2018ezm,MenciaUranga2018}, 
dense QCD of quark matter \cite{Eto:2021nle}, 
and the two-Higgs doublet model \cite{Eto:2021dca} 
as a model beyond the standard model 
of elementary particles. 
Compared with these systems, 
the unique feature of $^3P_2$ superfluids is that 
constituent half-quantized vortices are 
non-Abelian vortices, 
that is, characterized by a non-Abelian first homotopy group.

This paper is organized as follows.
In Sec.~\ref{sec:formulation}, we begin with  
formulations of $^3P_2$ superfluids within the GL 
approach in our notation.
Section \ref{sec:result} shows our numerical results for vortex states in the $^3P_2$ superfluids.
Section \ref{sec:summary} is devoted to a summary 
and discussion.





\section{Ginzburg-Landau free energy for 
$^3P_2$ neutron superfluids}\label{sec:formulation}

We start from a brief review of the GL theory for $^3P_2$
 superfluids \cite{Yasui:2019unp} 
 reformulated in the notation of Ref.~\cite{Kobayashi:2021arv}.
The effective GL Lagrangian density $f$ is given by 
\begin{align}
\begin{aligned}
f &= K_0 \left(f_{202}^{(0)} + f_{202}^{(1)}\right) + \alpha f_{002} + \beta_0 f_{004} + \gamma_0 f_{006} \\
&+ \delta_0 f_{008} + \beta_2 f_{022} + \gamma_2 f_{024} \\
&+ \sum_{4 l + 2 m + n = 10} O(\nabla^l |\Vec{B}|^m A^n),
\end{aligned}
\label{eq:Lagrangian}
\end{align}
where $f_{lmn}$ is the free energy part including 
$l$ spatial derivatives $\nabla$, 
$m$-th order of the magnetic field $\Vec{B}$, and $n$-th order of spin-2 spinor order parameter $\psi = (\psi_2, \psi_1, \psi_0, \psi_{-1}, \psi_{-2})^T$.
The spatial derivative term $f_{202}$ is further separated into current-spin independent and dependent parts $f_{202}^{(0)}$ and $f_{202}^{(1)}$, respectively.
Their specific forms can be written as
\begin{widetext}
\begin{align}
\begin{aligned}
&
f_{202}^{(0)} = 3 \Vec{j}^\dagger \cdot \Vec{j}, \\
&  f_{202}^{(1)} = 4 \Vec{j}^\dagger \cdot \Vec{j} - \frac{i}{2} \Vec{j}^\dagger \cdot \hat{\Vec{S}} \times \Vec{j} - \left( \Vec{j}^\dagger \cdot \hat{\Vec{S}} \right) \left( \hat{\Vec{S}} \cdot \Vec{j} \right), \\ &
f_{002} = 3 \rho, \\
&f_{004} = 6 \rho^2 + \frac{3}{4} \Vec{S}^2 - \frac{3}{2} |\Psi_{20}|^2, \\
&
f_{006} = - 324 \rho^3 - 81 \rho \Vec{S}^2 + 162 \rho |\Psi_{20}|^2 + 15 |\Psi_{30}|^2 - 27 |\Phi_{30}|^2, \\ &
f_{008} = 6480 \rho^4 + 1944 \rho^2 \Vec{S}^2 - 5184 \rho^2 |\Psi_{20}|^2 - 864 \rho |\Psi_{30}|^2 + 2592 \rho |\Phi_{30}|^2 + 81 \Vec{S}^4 + 648 |\Psi_{20}|^4 - 1296 \Gamma_{4} \\ &
f_{022} = 2 \rho \Vec{B}^2 - \frac{1}{2} \psi^\dagger \hat{S}_{\Vec{B}} \hat{S}_{\Vec{B}} \psi, \\ &
f_{024} = \left( -106 \rho^2 + \frac{9}{2} \Vec{S}^2 + 31 |\Psi_{20}|^2 \right) \Vec{B}^2 \\
&\phantom{f_{024}} + \left( 22 \rho \psi^\dagger \hat{S}_{\Vec{B}} \hat{S}_{\Vec{B}} \psi + \mathrm{Re}\left[ \Psi_{20}^\ast \psi^T \hat{S}_{\Vec{B}}^T \hat{T} \hat{S}_{\Vec{B}} \psi \right] \vphantom{\frac{5}{4}} + \frac{5}{4} \Psi_{22}^\dagger \hat{S}_{\Vec{B}} \hat{S}_{\Vec{B}} \Psi_{22} + \frac{1}{2} \Phi_{22}^T \hat{S}_{\Vec{B}}^T \hat{T} \hat{S}_{\Vec{B}} \Phi_{22} \right),
\end{aligned}
\end{align}
\end{widetext}
where $\hat{S}_i$ ($i = x,y,z$) are $5 \times 5$ spin-2 matrices, $\hat{T}$ is the time reversal operator defined by $(\hat{T} \psi)_m \equiv (-1)^m \psi_{-m}$, and $\hat{S}_{\Vec{B}} \equiv \Vec{\hat{S}} \cdot \Vec{B}$.
The $\psi$-dependent terms $\Vec{j}$, $\rho$, $\Vec{S}$, $\Psi_{20}$, $\Psi_{30}$, $\Phi_{30}$, $\Gamma_4$ are defined by 
\begin{align}
\begin{aligned}
& \Vec{j} = - i \nabla \psi, \quad
\rho = \psi^\dagger \psi, \quad
\Vec{S} = \psi^\dagger \hat{\Vec{S}} \psi, \\
& \Psi_{20} = \psi^T \hat{T} \psi = \sqrt{5} \sum_{m_1, m_2 = -2}^2 C^{00}_{2 m_1,2 m_2} \psi_{m_1} \psi_{m_2}, \\
& \Psi_{30}
= - \sqrt{\frac{35}{2}} \sum_{J=0}^4 \sum_{M = -J}^J \sum_{m_1, m_2, m_3 = -2}^2 \\
&\phantom{\Psi_{30}} \times C^{00}_{JM,2m_3} C^{JM}_{2m_1,2m_2} \psi_{m_1} \psi_{m_2} \psi_{m_3}, \\
& \Phi_{30}
= - \sqrt{\frac{35}{2}} \sum_{J=0}^4 \sum_{M = -J}^J \sum_{m_1, m_2, m_3 = -2}^2 \\
&\phantom{\Phi_{30}}\times C^{00}_{JM,2m_3} C^{JM}_{2m_1,2m_2} \psi_{m_1} \psi_{m_2} \varphi^\ast_{m_3}, \\
& \Gamma_4 = \mathrm{Re}\left[\Psi_{20} \Phi_{30}^{\ast 2}\right].
\end{aligned}
\end{align}
The GL coefficients can be obtained in the weak coupling limit within the quasiclassical approximation starting from the nonrelativistic spin-1/2 fermion field theory as \cite{Yasui:2019unp}
\begin{equation}
\begin{alignedat}{2}
& K_0 = \frac{7 \zeta(3) N(0) p_{\rm F}^4}{240 \pi^2 m_{\rm n}^2 T^2},\quad & & \alpha = \frac{N(0) p_{\rm F}^2}{3} \log\frac{T}{T_{\rm c}},\\
& \beta_0 = \frac{7 \zeta(3) N(0) p_{\rm F}^4}{60 \pi^2 T^2},\quad & & \gamma_0 = \frac{31 \zeta(5) N(0) p_{\rm F}^6}{13440 \pi^4 T^4},\\
& \delta_0 = \frac{127 \zeta(7) N(0) p_{\rm F}^8}{387072 \pi^6 T^6},\quad & & \beta_2 = \frac{7 \zeta(3) N(0) p_{\rm F}^2 \gamma_{\rm n}^2}{48 (1 + F_0^a)^2 \pi^2 T^2},\\
& \gamma_2 = \frac{31 \zeta(5) N(0) p_{\rm F}^4 \gamma_{\rm n}^2}{3840 (1 + F_0^a)^2 \pi^4 T^4}, & &
\end{alignedat}
\end{equation}
with the temperature $T$, the critical temperature $T_{\rm c}$, the neutron mass $m_{\rm n}$, the neutron gyromagnetic ratio $\gamma_{\rm n}$, the Fermi momentum $p_{\rm F}$, the state-number density $N(0) = m_{\rm n} p_{\rm F} / (2 \pi)^2$ at the Fermi surface, and the Landau parameter $F_0^a$.

The spin-2 spinor order parameter is often written by the $3\times 3$ traceless symmetric matrix $A$ given by
\begin{align}
\begin{aligned}
& [A]_{11} = \frac{\sqrt{3}}{2} (\psi_2 + \psi_{-2}) - \frac{1}{\sqrt{2}} \psi_0, \\
& [A]_{12} = [A]_{21} = \frac{\sqrt{3} i}{2} (\psi_2 - \psi_{-2}), \\
& [A]_{13} = [A]_{31} = - \frac{\sqrt{3}}{2} (\psi_1 - \psi_{-1}), \\
& [A]_{22} = - \frac{\sqrt{3}}{2} (\psi_2 + \psi_{-2}) - \frac{1}{\sqrt{2}} \psi_0, \\
& [A]_{23} = [A]_{32} = - \frac{\sqrt{3} i}{2} (\psi_1 + \psi_{-1}), \\
& [A]_{33} = \sqrt{2} \psi_0.
\end{aligned}
\label{eq:translation}
\end{align}

All candidates for uniform ground states were classified in 
Ref.~\cite{Kobayashi:2021arv}, and characterized by $U(1) \times SO(3)$ invariants $\Vec{S}^2$, $|\Psi_{20}|^2$, and $|\Psi_{30}|^2$.
The five characteristic symmetric states are ferromagnetic ($\Vec{S}^2 / \rho^2 = 4 \rho^2$, $|\Psi_{20}|^2 = |\Psi_{30}|^2 = 0$), uniaxial nematic (UN) ($\Vec{S}^2 = 0$, $|\Psi_{20}|^2 = \rho^2$, $|\Psi_{30}|^2 = \rho^3$), $D_4$ biaxial nematic (BN) ($\Vec{S}^2 = 0$, $|\Psi_{20}|^2 = \rho^2$, $|\Psi_{30}|^2 = 0$), $D_2$ BN ($\Vec{S}^2 = 0$, $|\Psi_{20}|^2 = \rho^2$, $0 < |\Psi_{30}|^2 < \rho^3$), and cyclic ($\Vec{S}^2 = 0$, $|\Psi_{20}|^2 = 0$, $|\Psi_{30}|^2 = 2 \rho^3$) states.

For the effective Lagrangian density $f$ in Eq.~\eqref{eq:Lagrangian}, the UN, $D_2$ BN, and $D_4$ BN states are predicted to be realized at $|\Vec{B}| = 0$, $0 < |\Vec{B}| < B_{\rm c}$, and $|\Vec{B} > B_c$, in $^3P_2$ superfluids 
\cite{Mizushima:2016fbn} as shown in Table \ref{table}(a). 
The critical magnetic field $B_{\rm c}$ separating the $D_2$ BN and $D_4$ BN states depends on the temperature and takes the maximum value $B_{\rm c} = 7.06 \times 10^{-2} \pi (1 + F_0^a) T_{\rm c} / \gamma_{\rm n}$ at $T \simeq 0.854 T_{\rm c}$.
With an estimation for the critical temperature $T_{\rm c} \approx 0.2$ MeV and the Landau parameter $F_0^a \approx 1$, this critical magnetic field can be estimated as $B_{\rm c} \approx 7.36 \times 10^{15}$ G.
At $T \lesssim 0.796 T_{\rm c}$, we obtain $B_{\rm c} = 0$.

\section{Vortex solutions}\label{sec:result}

\subsection{Ansatz}
Next, we consider vortex solutions with 
vortex cores placed at $r = 0$ in the cylindrical coordinate $(r,\theta,z)$ 
and the boundary $\psi|_{r \to \infty}$ far from vortex cores. 
The vortices and the angular momentum are parallel to the $z$ axis.
For the so-called integer vortices, the order parameters behave as $\psi_{-2 \leq m \leq 2}|_{r \to \infty} \propto e^{i \theta}$ around which the overall phase of $\psi$ winds by $2 \pi$.
In this section, we determine the boundary conditions 
around the vortices 
for the cases of $\Vec{B} = 0$ and $\Vec{B} \neq 0$,

For $\Vec{B} = 0$, the uniform ground state is degenerate within the possible UN state
\begin{align}
\begin{aligned}
&
\tilde{\psi}_{\pm 2} = \frac{e^{i (\phi \mp 2 a)} \sqrt{3} \sin^2 b}{2 \sqrt{2}}, \quad
\tilde{\psi}_{\pm 1} = \mp \frac{e^{i (\phi \mp a)} \sin(2 b)}{2 \sqrt{2}}, \\
&
\tilde{\psi}_0 =\frac{e^{i \phi} \{1 + 3 \cos(2 b)\}}{4}, 
\end{aligned}
\label{eq:uniaxial-nematic}
\end{align}
or
\begin{align}
\begin{aligned}
&\tilde{A} = \frac{1}{\sqrt{2}} R_z^a R_y^b \begin{pmatrix} -1 & 0 & 0 \\ 0 & -1 & 0 \\ 0 & 0 & 2 \end{pmatrix} R_y^{-b} R_z^{-a}, \\
& R_y^b \equiv \begin{pmatrix} \cos b & 0 & \sin b \\ 0 & 1 & 0 \\ -\sin b & 0 & \cos b \end{pmatrix}, \\
& R_z^a \equiv \begin{pmatrix} \cos a & -\sin a & 0 \\ \sin a & \cos a & 0 \\ 0 & 0 & 1 \end{pmatrix},
\end{aligned}
\end{align}
for $0 \leq a, 2 b, \phi \leq 2 \pi$.
Here $\tilde{\psi}$ is defined as $\tilde{\psi} \equiv \psi / \sqrt{\rho}$, and  $\tilde{A}$ is defined by replacing $A$ and $\psi$ with $\tilde{A}$ and $\tilde{\psi}$, respectively in Eq. \eqref{eq:translation}.
$a$, $b$, and $\phi$ represent overall spin rotations along $z$ axis and $y$ axis, and overall phase shift, respectively.
Under the spatial phase gradient $e^{i \theta}$ for the vortex solution, however, the current-spin-dependent free energy density $f_{202}^{(2)}$ in Eq.~\eqref{eq:Lagrangian} favors $b = 0$, giving
\begin{align}
\tilde{\psi} \xrightarrow{r \to \infty} \left(0,0,e^{i \theta},0,0\right)^T,
\label{eq:zero_field_solution}
\end{align}
or
\begin{align}
\tilde{A} \xrightarrow{r \to \infty} \frac{e^{i \theta}}{\sqrt{2}} \begin{pmatrix} -1 & 0 & 0 \\ 0 & -1 & 0 \\ 0 & 0 & 2 \end{pmatrix},
\end{align}

We next consider the case of $|\Vec{B}| > 0$.
In this paper, the simplest situation $\Vec{B} \parallel \Vec{z}$ is considered, i.e., the magnetic field is parallel to the angular momentum for the vortex.
For $0 < |\Vec{B}| < B_{\rm c}$, field-dependent free energy density $\beta_2 f_{022} + \gamma_2 f_{024}$ favors $\psi_{\pm 2}$, leading
\begin{align}
\tilde{\psi} \xrightarrow{r \to \infty} e^{i \theta} \left( \frac{e^{- 2 i a} \sin g}{\sqrt{2}}, 0, \cos g, 0, \frac{e^{2 i a} \sin g}{\sqrt{2}} \right)^T,
\label{eq:D2-biaxial_vortex}
\end{align}
or
\begin{align} 
\begin{aligned}
&\tilde{A} \xrightarrow{r \to \infty} \sqrt{2} e^{i \theta} R_z^a \begin{pmatrix} \sin g_- & 0 & 0 \\ 0 & -\sin g_+ & 0 \\ 0 & 0 & \cos g \end{pmatrix} R_z^{-a}, \\
& g_{\pm} \equiv g \pm \frac{\pi}{6},
\end{aligned}
\end{align}
where $g$ depends on $|\Vec{B}|$ and satisfies $\pi / 3 < g < \pi / 2$, making $\psi|_{r\to\infty}$ to be the $D_2$ BN state.
$a$ also represents the overall spin rotation along $z$ axis as well as that in Eq. \eqref{eq:uniaxial-nematic}.
Without current-spin dependent free energy $f_{202}^{(1)}$, $a$ takes the arbitrary value, but is fixed with the finite $f_{202}^{(1)}$ to minimize this.
In the limit of $|\Vec{B}| \searrow 0$, $g$ becomes $g \to \pi / 3$ giving
\begin{align}
\tilde{\psi} \xrightarrow{r \to \infty} e^{i \theta} \left( \frac{e^{- 2 i a} \sqrt{3}}{2 \sqrt{2}}, 0, \frac{1}{2}, 0, \frac{e^{2 i a} \sqrt{3}}{2 \sqrt{2}} \right)^T,
\label{eq:zero_limit_solution}
\end{align}
or
\begin{align} 
\tilde{A} \xrightarrow{r \to \infty} \frac{e^{i \theta}}{\sqrt{2}} R_z^a \begin{pmatrix} 1 & 0 & 0 \\ 0 & -2 & 0 \\ 0 & 0 & 1 \end{pmatrix} R_z^{-a}.
\end{align}
This solution belongs to the uniaxial nematic state in Eq. \eqref{eq:uniaxial-nematic} with $b = \pi/2$, but is different from that for $\Vec{B} = 0$ shown in Eq. \eqref{eq:zero_field_solution}, which leads the discontinuity between $\Vec{B} = 0$ and $|\Vec{B}| \searrow 0$.

In the limit of $|\Vec{B}| \nearrow B_{\rm c}$, $g$ becomes $g \to \pi/2$, giving 
\begin{align}
\tilde{\psi} \xrightarrow{r \to \infty} e^{i \theta} \left( \frac{e^{- 2 i a}}{\sqrt{2}}, 0, 0, 0, \frac{e^{2 i a}}{\sqrt{2}} \right)^T,
\label{eq:D4-biaxial_vortex}
\end{align}
or
\begin{align} 
\tilde{A} \xrightarrow{r \to \infty} \sqrt{\frac{3}{2}} e^{i \theta} R_z^a \begin{pmatrix} 1 & 0 & 0 \\ 0 & -1 & 0 \\ 0 & 0 & 0 \end{pmatrix} R_z^{-a}.
\end{align}
which belongs to the $D_4$ BN state.

For $|\Vec{B}| \geq B_{\rm c}$, the vortex state is the same as that in Eq.~\eqref{eq:D4-biaxial_vortex} that belongs to $D_4$ BN state.

For the half-quantized vortex, the overall phase of $\psi$ winds by $\pi$.
To keep the single-valued property of the order parameter, the spin also rotates.
Only the $D_4$ BN state under $|\Vec{B}| > B_{\rm c}$ enables the half-quantized vortex
with the order parameter giving
\begin{align}
\tilde{\psi} \xrightarrow{r \to \infty} \left( \frac{e^{- 2 i a}}{\sqrt{2}}, 0, 0, 0, \frac{e^{i (\theta + 2 a)}}{\sqrt{2}} \right)^T,
\label{eq:half-vortex-plus-ansatz}
\end{align}
or
\begin{align}
\begin{aligned}
& \tilde{A} \xrightarrow{r \to \infty} \frac{\sqrt{3}}{2 \sqrt{2}} R_z^a
\begin{pmatrix} \theta_+ & i \theta_- & 0 \\ i \theta_- & - \theta_+ & 0 \\ 0 & 0 & 0 \end{pmatrix} R_z^{-a}, \\
& \theta_{\pm} \equiv e^{i\theta} \pm 1
\end{aligned}
\end{align}
in the case of $\Vec{B} \parallel \Vec{z}$.
For the vortex solution \eqref{eq:half-vortex-plus-ansatz}, the $z$ component of the spin rotates by $\pi/2$ around the vortex.
The other solution with the spin rotation by $-\pi/2$ is
\begin{align}
\tilde{\psi} \xrightarrow{r \to \infty} \left( \frac{e^{i (\theta - 2 a)}}{\sqrt{2}}, 0, 0, 0, \frac{e^{2 i a}}{\sqrt{2}} \right)^T,
\label{eq:half-vortex-minus-ansatz}
\end{align}
or
\begin{align} 
\tilde{A} \xrightarrow{r \to \infty} \frac{\sqrt{3}}{2 \sqrt{2}} R_z^a
\begin{pmatrix} \theta_+ & - i \theta_- & 0 \\ - i \theta_- & - \theta_+ & 0 \\ 0 & 0 & 0 \end{pmatrix} R_z^{-a}.
\end{align}

\subsection{Numerical results} \label{subsec:result}

\begin{figure*}
\centering
\includegraphics[width=0.8\linewidth]{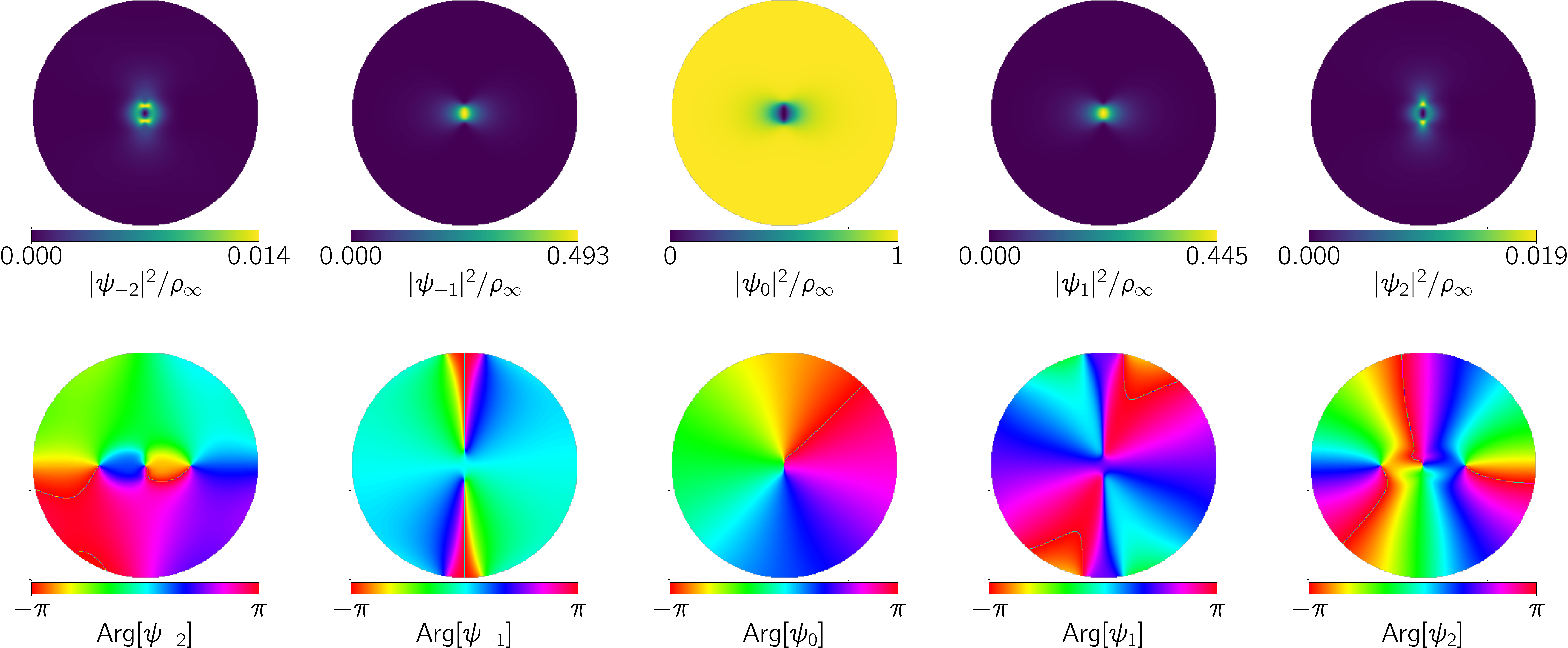}
\includegraphics[width=0.8\linewidth]{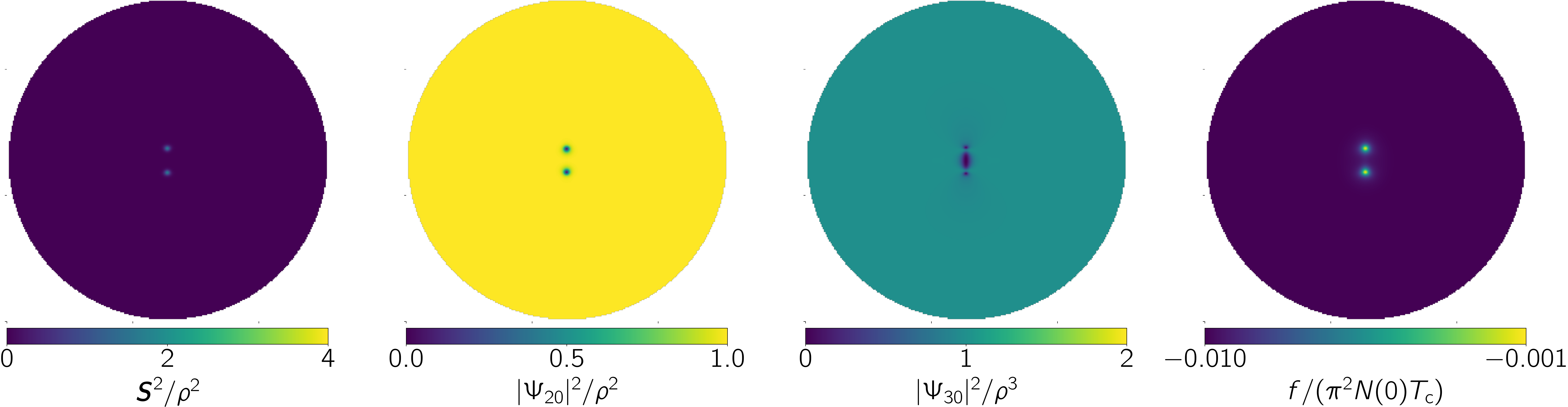}
\caption{
\label{fig:wave-000}
The vortex state 
in the UN phase at the zero magnetic field $\Vec{B} = 0$.
The squared modulus $|\psi_m|^2$ (top row), argument $\mathrm{Arg}[\psi_m]$ (middle row) of the order parameter,  
the $U(1) \times SO(3)$ invariants $\Vec{S}^2$, $|\Psi_{20}|^2$, and $|\Psi_{30}|^2$, and the free-energy density $f$ (bottom row) 
are shown. 
The radius of the figure shown here is $64 p_{\rm F} / (\pi m_{\rm n} T_{\rm c}) \approx 1.33 \times 10^4$ fm.
The two half-quantized vortices are connected by a single line soliton 
of the $D_4$ BN order.
}
\end{figure*}
\begin{figure*}
\centering
\includegraphics[width=0.8\linewidth]{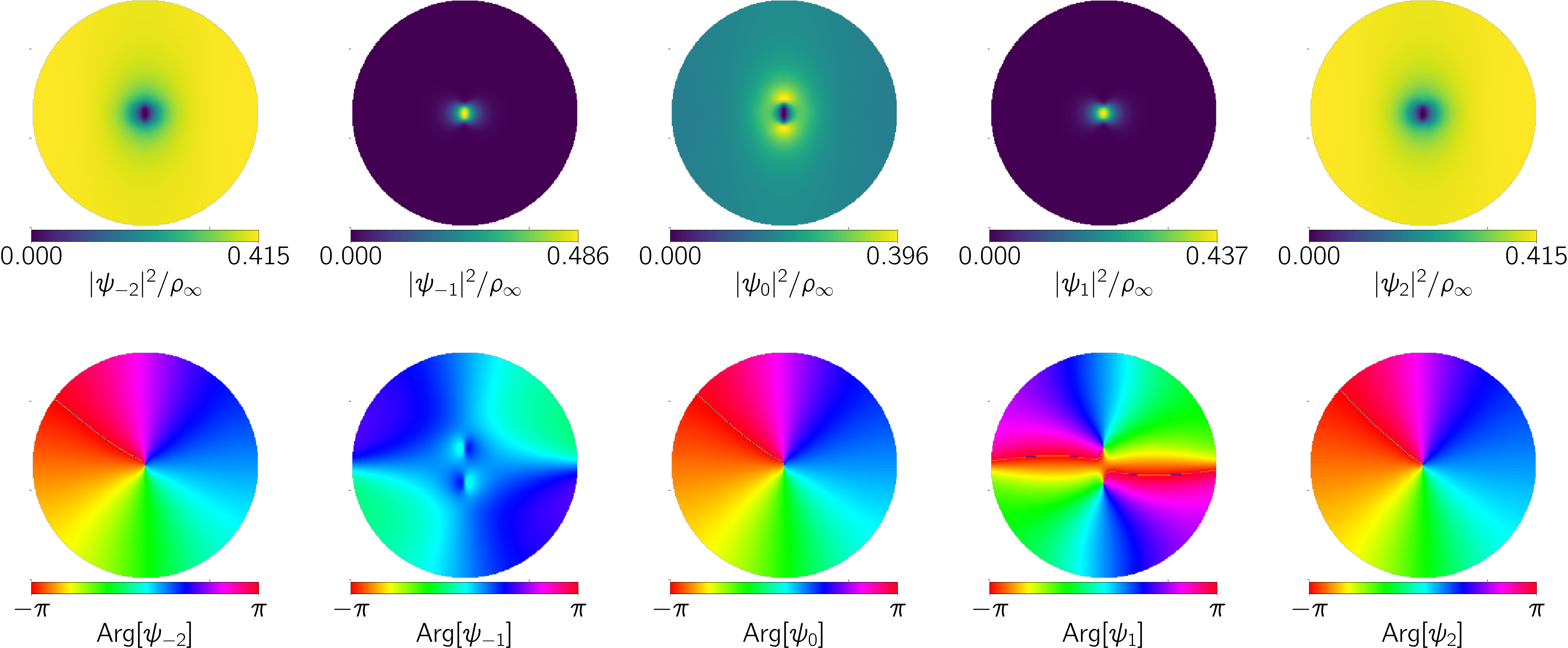}
\includegraphics[width=0.8\linewidth]{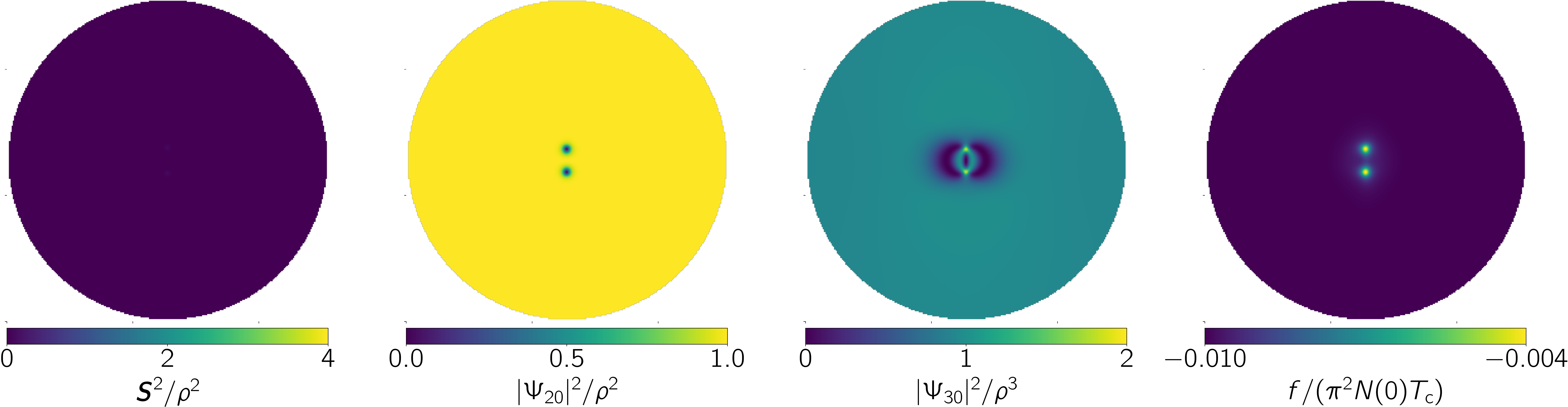}
\caption{
\label{fig:wave-050}
The vortex state in the $D_2$ BN phase 
at the magnetic field $\Vec{B} = 0.5 B_{\rm c} \hat{\Vec{z}}$.
The squared modulus $|\psi_m|^2$ (top row), argument $\mathrm{Arg}[\psi_m]$ (middle row) of the order parameter, the $U(1) \times SO(3)$ invariants $\Vec{S}^2$, $|\Psi_{20}|^2$, and $|\Psi_{30}|^2$, and the free-energy density $f$ (bottom row) are shown. 
The radius of the figure shown here is $64 p_{\rm F} / (\pi m_{\rm n} T_{\rm c}) \approx 1.33 \times 10^4$ fm.
The two half-quantized vortices are connected by 
three line solitons with 
the $D_4$ BN order in their cores.
}
\end{figure*}

In this subsection, we show the numerical results for the overall vortex state by minimizing the free-energy $f$ under the boundary condition
\begin{align}
\psi_m(\theta+\pi/2) = i \psi_m(\theta),
\label{eq:integer-vortex_boundary}
\end{align}
or
\begin{align}
[A]_{ij}(\theta+\pi/2) = i [A]_{ij}(\theta),
\end{align}
at the boundary $r = L/2$, which induces the integer vortex solution.
The minimization of the free energy density $f$ can be done by finding the solution of the stationary solution of the GL equation
\begin{align}
\frac{\delta f}{\delta \psi_m^\ast} = 0.
\label{eq:GL_stationary}
\end{align}
The solution of Eq. \eqref{eq:GL_stationary} can be obtained by introducing the relaxation time $t$ and the dependence of the order parameter $\psi_m$ on $t$, and solving
\begin{align}
\dot{\psi}_m = - \frac{\delta f}{\delta \psi_m^\ast}.
\label{eq:GL_time_evolution}
\end{align}
After the long time evolution of Eq. \eqref{eq:GL_time_evolution}, we attain the solution of Eq. \eqref{eq:GL_stationary}.
With an appropriate scaling of the time $t$, Eq. \eqref{eq:GL_time_evolution} is nothing but the time-dependent GL equation that is often used in the research field of superconductivity.
However, it has not yet derived from the microscopic theory for the $^3P_2$ superfluids.

Here, $L$ is the system size fixed to be $L = 128 p_{\rm F} / [\pi^2 N(0) T_{\rm c}^2]$.
The temperature is fixed to be $T = 0.854 T_{\rm c} \approx 0.171$ MeV 
for which $B_{\rm c}$ takes the maximal value $B_{\rm c} = 7.06 \times 10^{-2} \pi (1 + F_0^a) T_{\rm c} / \gamma_{\rm n} \approx 7.36 \times 10^{15}$ G with the critical temperature $T_{\rm c} \approx 0.2$ MeV and Landau parameter $F_0^a \approx 1$.
The magnetic field $\Vec{B} (\parallel \Vec{z})$ changes from 0 to $1.5 B_{\rm c}$.

\subsubsection{The vortex state 
with ferromagnetic core in the UN phase} \label{subsubsec:ferrocore}
We start from the case of the UN phase at zero magnetic field.
Figure \ref{fig:wave-000} shows the squared modulus $|\psi_m|^2$, argument $\mathrm{Arg}[\psi_m]$ of the order parameter, $U(1) \times SO(3)$ invariants $\Vec{S}^2$, $|\Psi_{20}|^2$, and $|\Psi_{30}|^2$, and the free energy density $f$ at the zero magnetic field $\Vec{B} = 0$.
The radius of circles in figures is fixed to be $64 \hbar^2 k_{\rm F} / (\pi M_{\rm n}^\ast T_{\rm c}) \approx 1.33 \times 10^4$ fm.
Where $k_{\rm F} = (3 \pi^2 n)^{1/3} \approx 2.20$ fm$^{-1}$ is the neutron Fermi wave number with the neutron number density $n \approx 2.25 n_0 \approx 0.36$ fm$^{-3}$ for the saturation density $n_0 \approx 0.16$ fm$^{-3}$ of nuclear matter.
$M_{\rm n}^\ast$ is the effective neutron mass $M_{\rm n}^\ast c^2 \approx 0.7 M_{\rm n} c^2 \approx 658$ MeV for the neutron vacuum mass $M_{\rm n} c^2 \approx 940$ MeV.
The critical temperature is set to be $T_{\rm c} \approx 0.2$ MeV. 
The order parameter $\psi$ satisfies Eq.~\eqref{eq:zero_field_solution} near the edge of the system, where the state belongs to the uniaxial nematic state $\Vec{S}^2 = 0$, $|\Psi_{20}|^2 / \rho^2 = |\psi_{30}|^2 / \rho^3 = 1$.
At the center of the system, there are two holes of $|\Psi_{20}|^2$ implying the breakdown of the UN order. 
These two holes correspond to the vortex cores 
and each of them carries half circulations.
This result clearly shows that a singly quantized vortex splits into two half-quantized non-Abelian vortices around the vortex core.
The fact that isolated half-quantized vortices can topologically exist only in the $D_4$ BN state implies that 
the $D_4$ BN order should appear around the vortex core.
In fact, we can confirm that the $D_4$ BN order characterized by $\Vec{S}^2 = 0$, $|\Psi_{20}|^2 / \rho^2 = 1$, and $|\Psi_{30}|^2 / \rho^3 = 0$ appears along a line structure bridging two vortex cores, as can be seen in the plot of $|\Psi_{30}|^2$ locally inducing 
the $D_4$ BN order and half-quantized vortices.
On the other hand, 
at the vortex cores, the $U(1) \times SO(3)$ invariants are $\Vec{S}^2 /\rho^2 \simeq 4$ and $|\Psi_{20}|^2 / \rho^2 = |\psi_{30}|^2 / \rho^3 = 0$, implying the appearance of the ferromagnetic order. 
Therefore, we characterize this vortex by the ferromagnetic core.

In this phase, there is also a metastable vortex molecule 
state with the cyclic cores 
similar to the $D_2$ phase discussed below.

\subsubsection{The vortex state with the cyclic core in the $D_2$ BN phase}

At $|\Vec{B}| = 0.5 B_{\rm c}$ as shown in Fig. \ref{fig:wave-050}, the order parameter drastically changes from that in Eq. \eqref{eq:zero_field_solution} to that in Eq. \eqref{eq:D2-biaxial_vortex} where $\psi_{\pm 2}$ become finite at $r \to \infty$.
As well as the case for $\Vec{B} = 0$, a singly quantized vortex splits into two half-quantized vortices with two holes of $|\Psi_{20}|^2$.
In contrast to the vortex molecule in the UN phase, 
there are three soliton lines of $D_4$ BN order with $|\Psi_{30}|^2 / \rho^3 = 0$ bridging two half-quantized vortices. 
At the vortex cores, we have $\Vec{S}^2 /\rho^2 \simeq 0$ and $|\psi_{30}|^2 / \rho^3 \simeq 2$ supporting the cyclic order.
We characterize this vortex by the cyclic core. 
With turning off the magnetic field, it becomes 
a metastable state with higher energy than 
the lowest energy state of 
the vortex molecule discussed in Sec \ref{sec:result} \ref{subsec:result} \ref{subsubsec:ferrocore}.

\begin{figure*}
\centering
\includegraphics[width=0.8\linewidth]{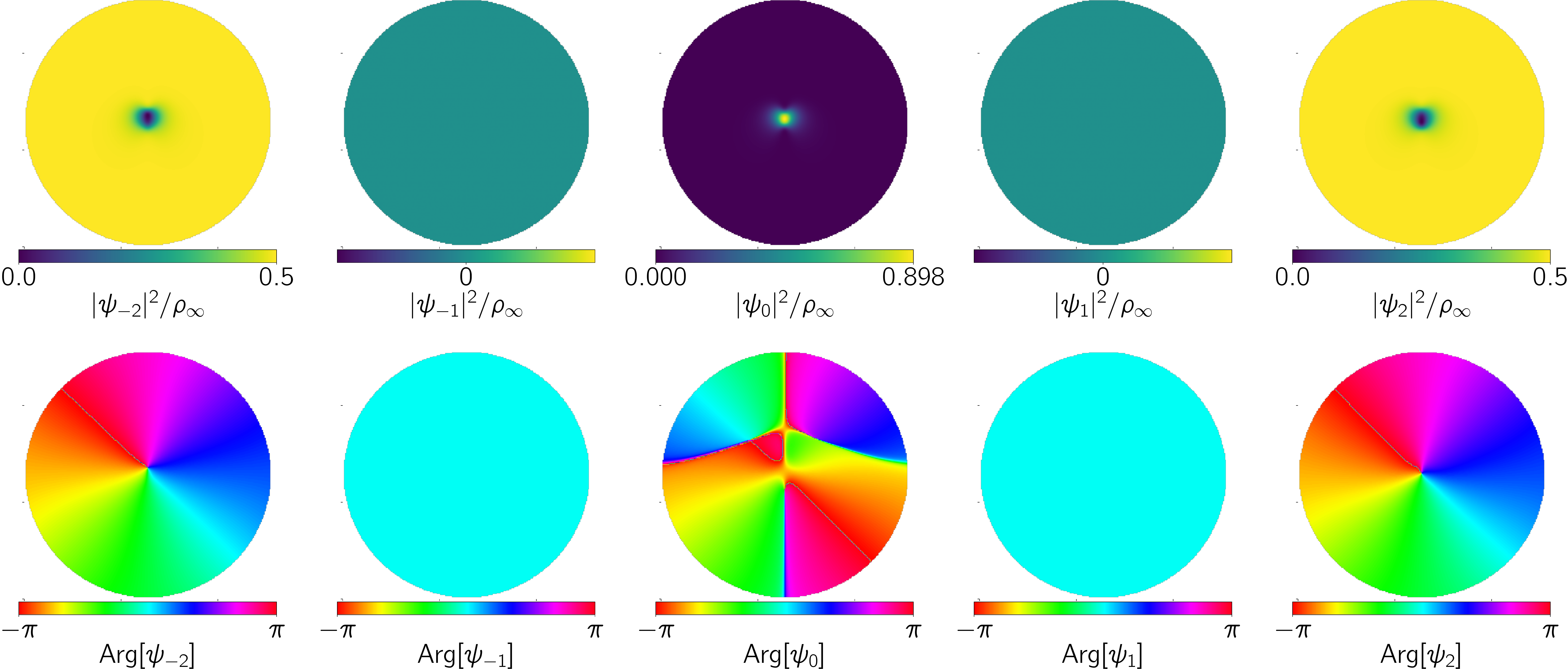}
\includegraphics[width=0.8\linewidth]{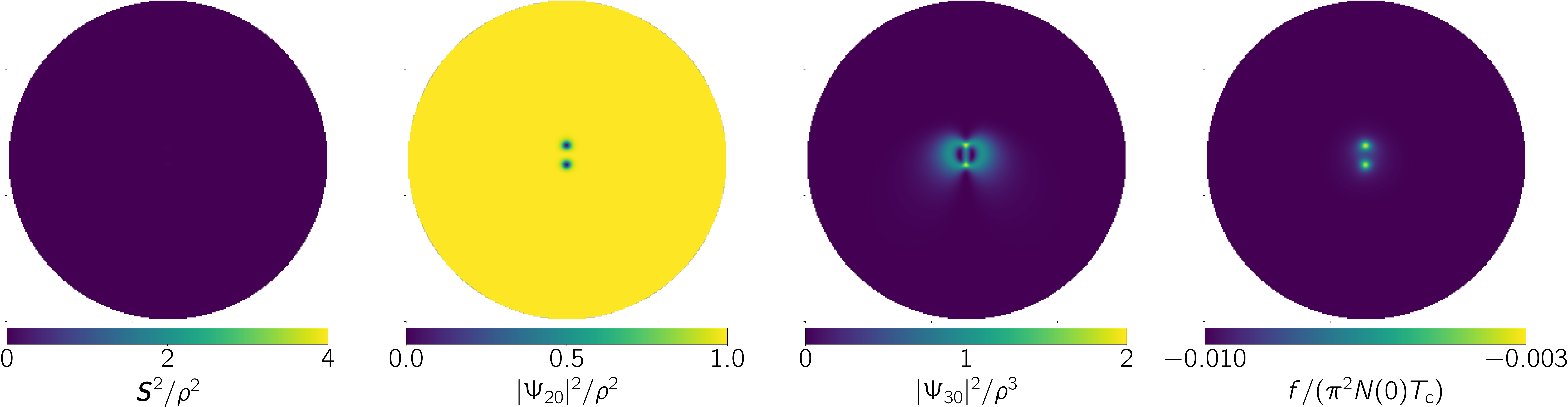}
\caption{
\label{fig:wave-130}
The vortex state in the $D_4$ BN phase 
at the magnetic field $\Vec{B} = 1.3 B_{\rm c} \hat{\Vec{z}}$.
Squared modulus $|\psi_m|^2$ (top row), argument $\mathrm{Arg}[\psi_m]$ (middle row) of the order parameter, the $U(1) \times SO(3)$ invariants $\Vec{S}^2$, $|\Psi_{20}|^2$, and $|\Psi_{30}|^2$, and the free-energy density $f$ are shown.
The radius of the figure shown here is $64 p_{\rm F} / (\pi m_{\rm n} T_{\rm c}) \approx 1.33 \times 10^4$ fm.
The two half-quantized vortices are connected by 
three line solitons of the $D_2$ BN order in their cores.
}
\centering
\includegraphics[width=0.8\linewidth]{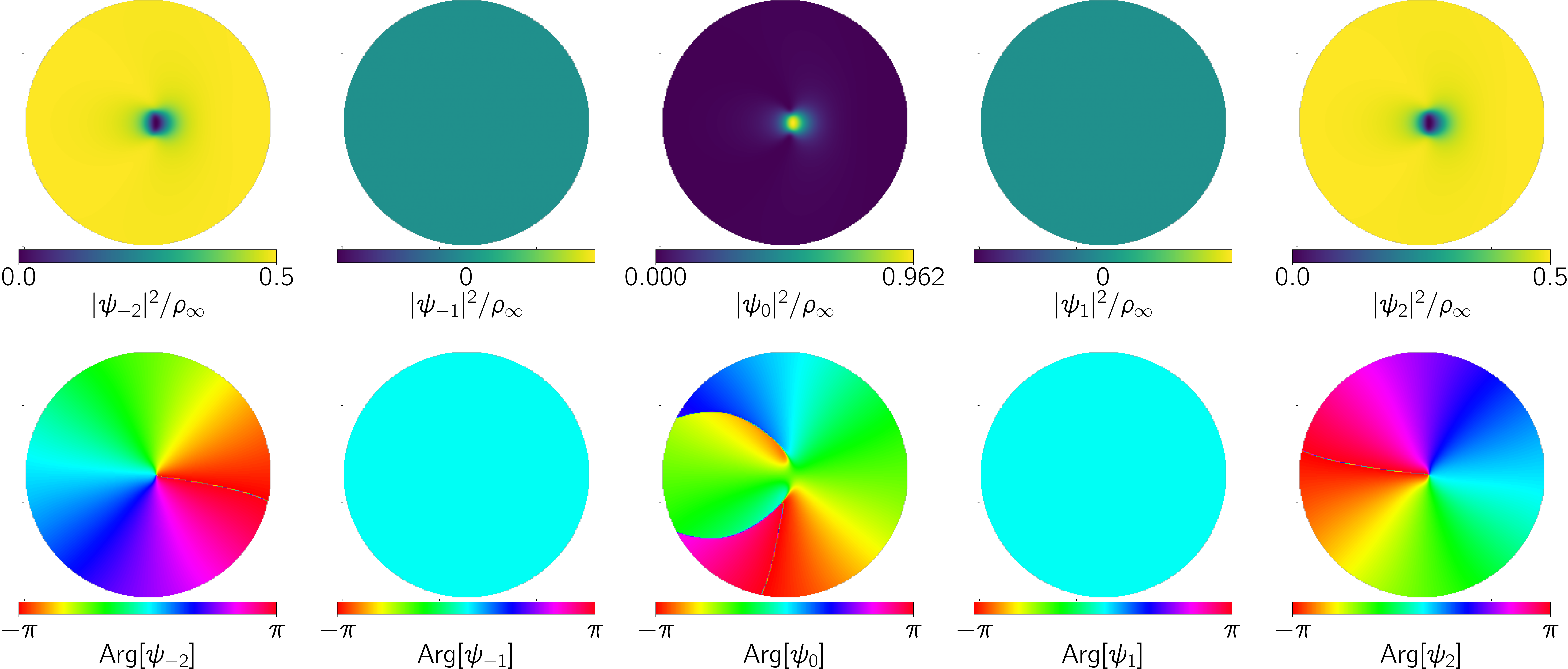}
\includegraphics[width=0.8\linewidth]{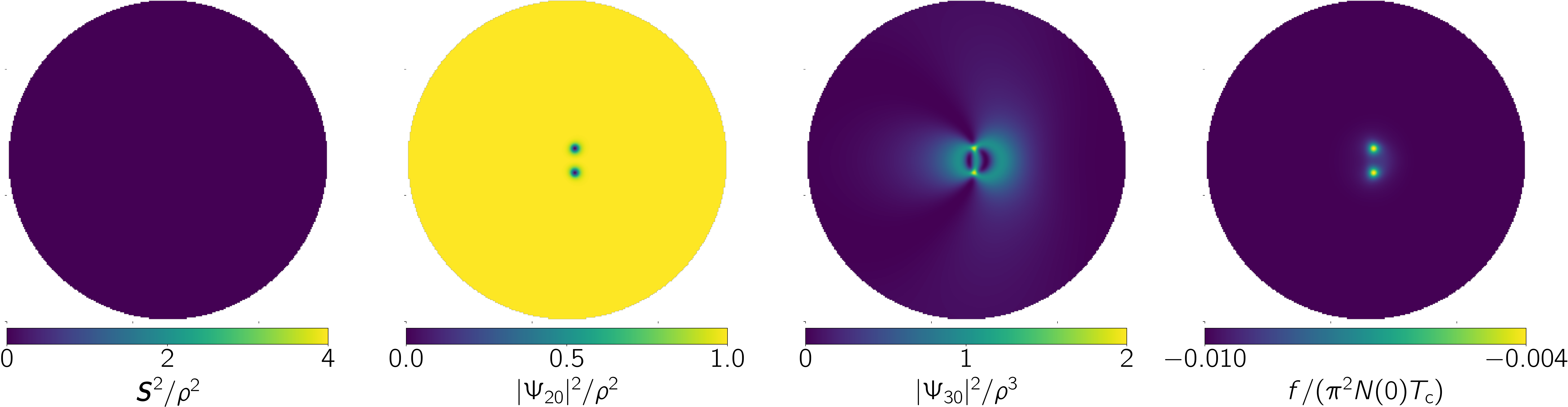}
\caption{
\label{fig:wave-100}
The vortex state at the boundary between 
the $D_2$ BN and $D_4$ BN phases  
 at the magnetic field $\Vec{B} = B_{\rm c} \hat{\Vec{z}}$.
The squared modulus $|\psi_m|^2$ (top row), argument $\mathrm{Arg}[\psi_m]$ (middle row) of the order parameter, the $U(1) \times SO(3)$ invariants $\Vec{S}^2$, $|\Psi_{20}|^2$, and $|\Psi_{30}|^2$, and the free-energy density $f$ are shown.
The radius of the figure shown here is $64 p_{\rm F} / (\pi m_{\rm n} T_{\rm c}) \approx 1.33 \times 10^4$ fm.
The two half-quantized non-Abelian vortices are connected by 
three line solitons asymmetrically.
}
\end{figure*}
\subsubsection{The vortex state with the cyclic core in the $D_4$ BN phase}	
At $|\Vec{B}| > B_{\rm c}$, the order parameter in the bulk 
becomes the $D_4$ BN state shown in Eq.~\eqref{eq:D4-biaxial_vortex} at $r \to \infty$.
Figure \ref{fig:wave-130} shows the $SO(3)$ invariants and the free-energy density at the magnetic field $|\Vec{B}| = 1.3 B_{\rm c}$. 
Although one half-quantized vortex is topologically stable in this state, two half-quantized vortices form 
a bound state to be an integer vortex.
The cores of half-quantized vortices are filled with the cyclic order having $\Vec{S}^2 /\rho^2 \simeq 0$ and $|\psi_{30}|^2 / \rho^3 \simeq 2$ as well as the case of the $D_2$ BN phase 
($0 < |\Vec{B}| < B_{\rm c}$).
In contrast to the case of the $D_2$ BN phase, the 
three line solitons 
bridging two half-quantized vortices are 
characterized by the $D_2$ BN order with 
$0 < |\Psi_{30}|^2 / \rho^3 < 1$. 
 
Our results suggest that the order of three line solitons 
connecting the two half-quantized vortices 
are exchanged between $D_2$ and $D_4$ BN orders for $|\Vec{B}| < B_{\rm c}$ and $|\Vec{B}| > B_{\rm c}$, respectively.
Another characteristic feature of the case of $|\Vec{B}| > B_{\rm c}$ 
is a fact that $\psi_{\pm 1}$ components completely vanish.

\subsubsection{The vortex state at the boundary between 
the $D_2$ and $D_4$ BN phases}

Figure \ref{fig:wave-100} shows the vortex state at the critical magnetic field $|\Vec{B}| = B_{\rm c}$.
Especially, we can see the asymmetric shape of $|\Psi_{30}|^2 / \rho^3$ as an intermediate state between those below and above the critical magnetic field $B_{\rm c}$, in which two $D_4$ and $D_2$ BN orders compete as candidates of 
the line solitons connecting the two half-quantized vortices. 
This asymmetric structure of the vortex core soon vanishes as the magnetic field $\Vec{B}$ becomes away from the critical magnetic field $B_{\rm c}$.
In our case, the vortex core becomes symmetric at $|\Vec{B}| = 0.9 B_{\rm c}$ and $|\Vec{B}| = 1.1 B_{\rm c}$.

\subsubsection{Energetics and distance between half-quantized vortices}
\begin{figure}
\centering
\includegraphics[width=0.99\linewidth]{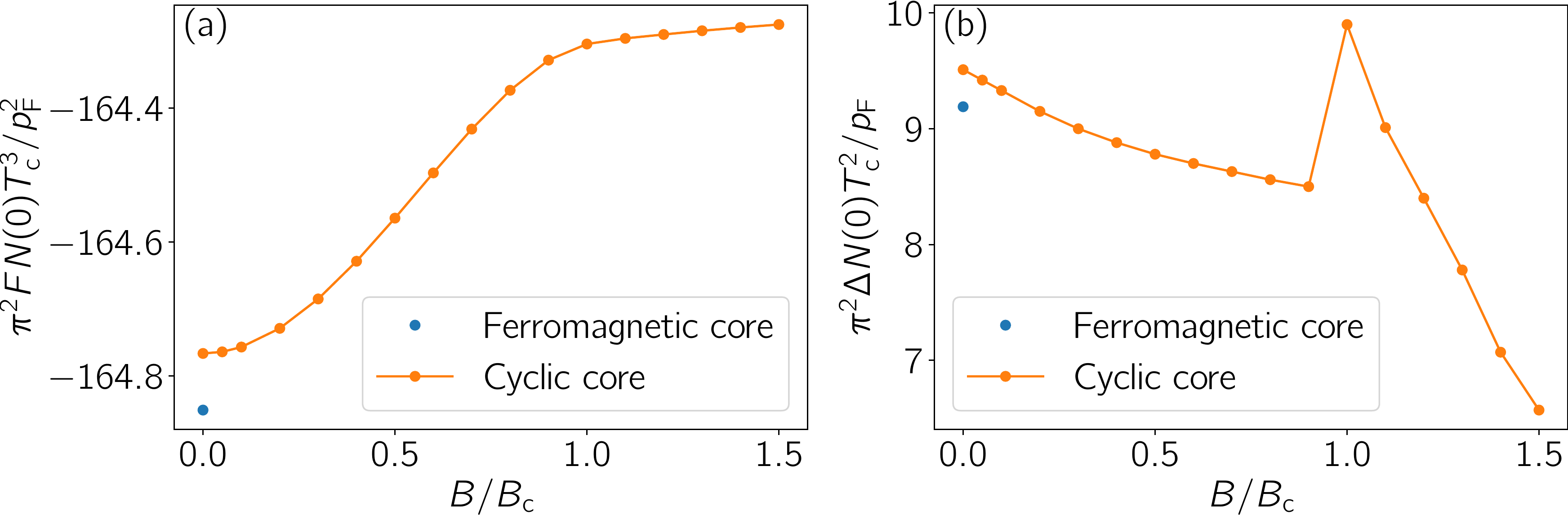}
\caption{
\label{fig:energy}
(a): Free energy $F$ and (b): distance $\Delta$ between two half-quantized  non-Abelian vortices as a function of the magnetic field $|\Vec{B}|$.
}
\end{figure}
Figure \ref{fig:energy} (a) shows the free energy $F = \int dr d\theta\: r f$ as a function of the magnetic field $|\Vec{B}|$.
The free energy $F$ monotonically 
increases with magnetic field, 
and it is continuous when the magnetic field across the critical one $B_{\rm c}$.
At the zero magnetic field $\Vec{B} = 0$, the vortex solution with 
ferromagnetic cores with the boundary condition in 
Eq.~\eqref{eq:zero_field_solution} 
has the lower free energy $F$ than that with the cyclic cores and the boundary shown in Eq.~\eqref{eq:D2-biaxial_vortex}, and discontinuously connects at $\Vec{B} = 0$.
However, a vortex solution with cyclic cores 
with the boundary in Eq.~\eqref{eq:D2-biaxial_vortex}  
can also exist as the metastable solution, 
continuously connecting to the solutions in the $D_2$ BN phase 
$|\Vec{B}| > 0$.

Figure \ref{fig:energy} (b) shows the distance $\Delta$ between two half-quantized vortices 
 as a function of the magnetic field $|\Vec{B}|$.
$\Delta$ monotonically decreases with the magnetic field $|\Vec{B}|$, discontinuously increases at $|\Vec{B}| = B_{\rm c}$, 
and again monotonically  decreases for $|\Vec{B}| > B_{\rm c}$. 
At the zero magnetic field $\Vec{B} = 0$, the half-quantized vortices having the ferromagnetic cores have smaller $\Delta$ than that for those having the cyclic cores.

\begin{figure*}
\centering
\includegraphics[width=0.8\linewidth]{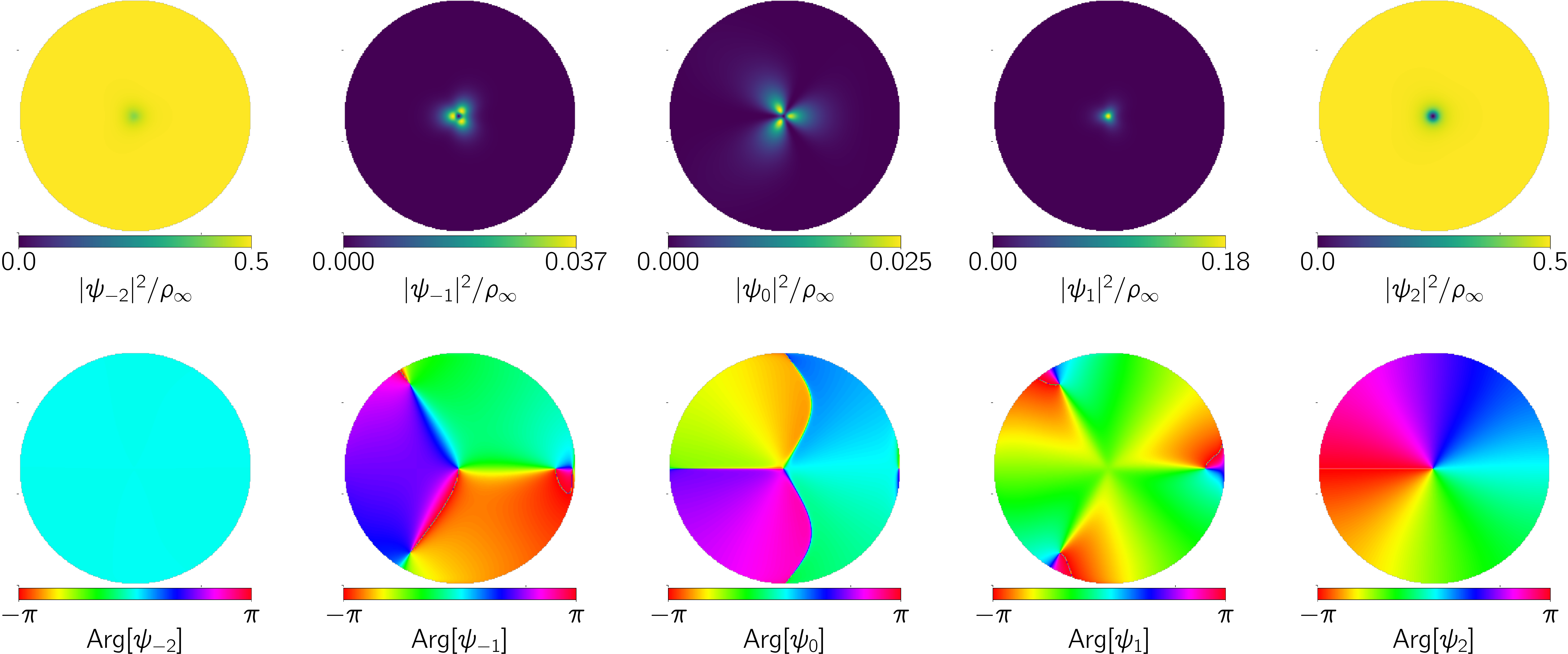}
\includegraphics[width=0.8\linewidth]{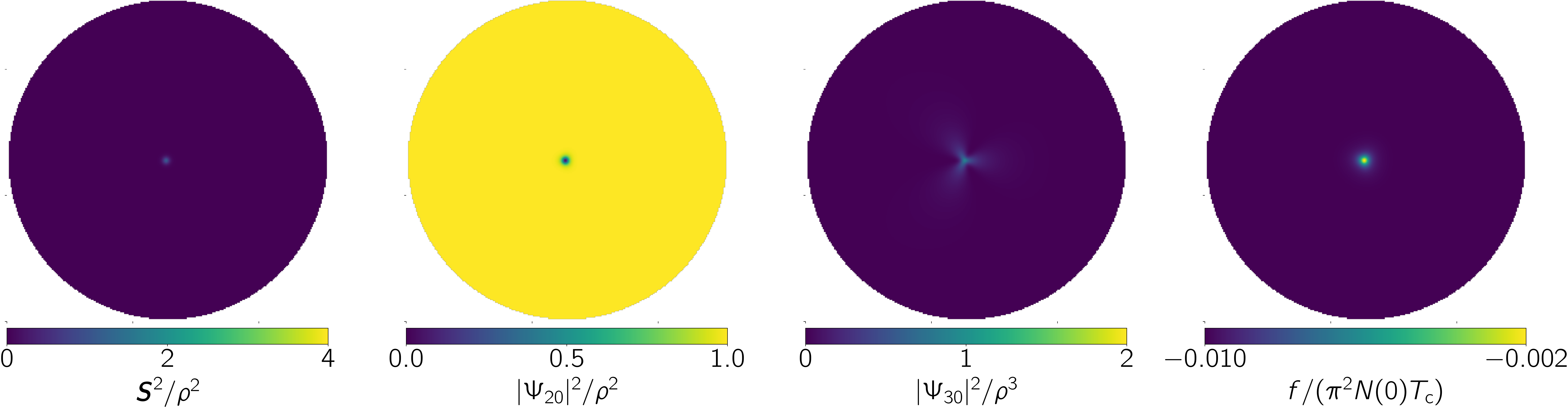}
\caption{
\label{fig:hwave-130}
A single isolated half-quantized non-Abelian vortex  
in the $D_4$ BN phase at the magnetic field $\Vec{B} = 1.3 B_{\rm c} \hat{\Vec{z}}$.
The squared modulus $|\psi_m|^2$ (top row), argument $\mathrm{Arg}[\psi_m]$ (middle row) of the order parameter, the $U(1) \times SO(3)$ invariants $\Vec{S}^2$, $|\Psi_{20}|^2$, and $|\Psi_{30}|^2$, and the free-energy density $f$ are shown. 
The radius of the figure shown here is $64 p_{\rm F} / (\pi m_{\rm n} T_{\rm c}) \approx 1.33 \times 10^4$ fm.
}
\end{figure*}

\subsubsection{Isolated half-quantized vortices 
in the $D_4$ BN phase}

Finally, we also obtain the solution for a single half-quantized vortex at $|\Vec{B}| > B_{\rm c}$. 
We put the boundary condition as
\begin{align}
\begin{aligned}
& \psi_2(\theta+\pi/2) = i \psi_2(\theta), \\
& \psi_{1,0,-1} = 0, \\
& \psi_{-2}(\theta+\pi) = \psi_{-2}(\theta),
\end{aligned}
\label{eq:half-vortex_boundary}
\end{align}
or
\begin{align}
\begin{aligned}
&{[A]}_{11}(\theta+\pi/2) = - [A]_{22}(\theta+\pi/2) \\
&\phantom{{[A]}_{11}(\theta+\pi/2)} = \frac{e^{i \pi / 4}}{\sqrt{2}} \left\{[A]_{11}(\theta) - [A]_{12}(\theta)\right\}, \\
&[A]_{12}(\theta+\pi/2) = [A]_{21}(\theta+\pi/2) \\
&\phantom{{[A]}_{12}(\theta+\pi/2)}= \frac{e^{i \pi / 4}}{\sqrt{2}} \left\{[A]_{11}(\theta) + [A]_{12}(\theta)\right\}, \\
&[A]_{13} = [A]_{23} = [A]_{31} = [A]_{32} = [A]_{33} = 0 \\
\end{aligned}
\end{align}
at the boundary which induces half-quantized vortex solutions.
Figure \ref{fig:hwave-130} shows a single half-quantized vortex state at $|\Vec{B}| = 1.3 B_{\rm c}$.
The system has a three-fold rotational symmetry around the vortex core which can be seen in $|\psi_{-1}|^2$, $|\psi_{0}|^2$, and $|\Psi_{30}|^2$.
The existence of the three-fold symmetry in $|\Psi_{30}|^2$ is closely related to the existence of the three soliton lines between two half-quantized vortices for the vortex shown in 
Fig.~\ref{fig:wave-130}.
Such a three-fold symmetry at the vortex core has been also observed in ultracold spin-2 atomic BECs \cite{Kobayashi:2011_arXiv}.

One of the main differences of the isolated half-quantized vortex in 
Fig.~\ref{fig:hwave-130} from the constituent one in the molecule 
in Fig.~\ref{fig:wave-130} is that the vortex core 
in Fig.~\ref{fig:hwave-130} 
is filled with the state having $\Vec{S}^2 / \rho^2 > 0$ and $|\Psi_{30}|^2 / \rho^3 > 0$, which corresponds to the mixed state \cite{Kobayashi:2021arv} being intermediate state between the ferromagnetic and cyclic states. 
Another difference is 
that $\psi_{\pm 1}$ takes finite values near the vortex core.

\section{Summary and Discussion}
\label{sec:summary}

In this paper, we have presented 
vortex solutions, that is, singly quantized vortices and 
half-quantized non-Abelian vortices,   
in the neutron $^3P_2$ superfluids in the case that the external magnetic field is parallel to the angular momentum of the vortices. 
We have found that 
a singly quantized vortex splits into two half-quantized non-Abelian vortices connected by soliton(s) 
forming a vortex molecule, 
at any strength of the magnetic field. 
The main results are summarized in Table \ref{table}.
In the absence of the magnetic field 
in which the UN state is the bulk ground state, 
the cores of the half-quantized vortices are filled with the ferromagnetic states, 
and a single linear soliton with the $D_4$ BN state connects 
the two half-quantized vortices, 
as shown in Fig.~\ref{fig:wave-000}.
At the finite magnetic field below the critical one 
separating $D_2$ and $D_4$ BN states, 
the bulk ground state is the $D_2$ BN state.
In this case,  
the cores of the half-quantized vortices are filled with the cyclic state 
and are connected by three line solitons composed of the $D_4$ BN order, 
as shown in Fig.~\ref{fig:wave-050}. 
Above the critical magnetic field for which 
the bulk ground state is the $D_4$ BN state, 
a single half-quantized vortex is topologically allowed stably, 
as constructed in Fig.~\ref{fig:hwave-130}. 
Nevertheless, two half-quantized vortices are confined 
with three line solitons composed of $D_2$ BN order, 
still forming a vortex molecule as shown in Fig.~\ref{fig:wave-130}. 
At the critical magnetic field, the vortex core becomes asymmetric 
as in Fig.~\ref{fig:wave-100}, 
as an intermediate state between the two kinds of molecules of two half-quantized vortices connected by the three $D_4$ BN solitons 
and those connected by the three $D_2$ BN solitons.
We have also found that 
the energy of the vortex molecule monotonically increases 
as the magnetic field increases, 
which is continuous including the critical magnetic field 
as in Fig.~\ref{fig:energy}(a).
The distance between the two half-quantized vortices decreases 
as the magnetic field increases,  
except for a discontinuous jump with an increase 
at the critical magnetic field in Fig.~\ref{fig:energy}(b).

Except for the case of the zero magnetic field, vortex cores are always filled with the cyclic state, which also appears even at the zero magnetic field as a metastable state as in Fig.~\ref{fig:energy}(a).
Our results contradict with the BdG approach \cite{Masaki:2019rsz,Masaki:2021hmk,Masaki:2022preparation} in which vortex cores are filled with the mixed state for $|\Vec{B}| < B_{\rm c}$ \cite{Masaki:2022preparation} without $\psi_{\pm 1}$ components and completely separated as two isolated half-quantized vortices without linear soliton \cite{Masaki:2021hmk} for $|\Vec{B}| > B_{\rm c}$.
A main possible reason for this contradiction comes from the difference of the GL and BdG approaches.
The latter approach has an advantage in describing the microscopic structure such as vortex cores and fermion degrees of freedom, and the cyclic state inside cores that we have obtained could be an artifact of the low energy theory.
We should study this point in more detail with, for example, 
a GL expansion to higher order.
Another minor possible reason is the difference of treatments of the boundary condition and positions of half-quantized vortices.
In the previous study, the boundary is fixed with the bulk integer vortex state [Eqs.~\eqref{eq:D2-biaxial_vortex} and \eqref{eq:D4-biaxial_vortex}].
The positions of the half-quantized vortices are also fixed and treated as a differential parameter.
On the other hand, neither boundary state determined in the boundary condition \eqref{eq:integer-vortex_boundary} nor the positions of half-quantized vortices are fixed and automatically determined to minimize the whole free-energy density in our study.
This subtle difference may affect the difference of the vortex-core states.

Here we address further discussions for future studies.
We here have studied only the case for the magnetic field parallel to the angular momentum (the direction of vortices). 
We will report the case for an arbitrary angle between them 
elsewhere, 
which should be important for study of neutron stars 
in more general situations.

Further studies should be done 
for multiple vortex states such as a vortex lattice 
under rapid rotation 
relevant for neutron star interiors.
In particular, it is important to study 
whether,  in a vortex lattice, 
 constituent half-quantized vortices 
are still tightly bound 
as 
a singly quantized vortex as found in this paper 
or they are separated by distances of the same order 
as two-component BECs \cite{
Mueller2002,Kasamatsu2003,Cipriani:2013nya} 
and whether a lattice is triangular or square.

Another important subject is 
a collision dynamics of vortices in three spatial dimensions.  
It is important whether two vortices reconnect in collision 
or  
a formation of a rung between them occurs   
as the case of non-Abelian vortices in 
the cyclic phase of spin-2 spinor BECs 
\cite{Kobayashi:2008pk}.
The presence or absence of such a vortex reconnection 
is crucial for states of the quantum turbulence.
With this regards, vortex reconnection was reported 
in the nematic phase of a spin-2 BEC \cite{Borgh:2016cco},  
a superfluid similar to $^3P_2$ superfluids.
Collision of two vortex molecules may be accompanied 
by swapping partners as found in 
vortex molecules in two-component BECs 
\cite{Eto:2019uhe}.

The coexistence of $^1S_0$ and $^3P_2$ superfluids 
drastically changes the phase diagram 
\cite{Yasui:2020xqb}, 
and vortex states in this case will be also one direction to be 
explored. 
In this case, vortices having 
winding only in either of $^1S_0$ and $^3P_2$ condensates 
would be further fractionalized. 
In addition,
vortex states in the ferromagnetic phase 
appearing without quasi-classical approximation in the region 
close to the critical temperature 
\cite{Mizushima:2021qrz} are also interesting 
to be investigated.

Recently, 
it has been proposed that 
in the deep inside of neutron star cores, 
the quark-hadron continuity for two-flavor quarks holds; 
the $^3P_2$ superfluid (nuclear matter) 
is continuously connected through crossover to 
a two-flavor quark matter called the 2SC+$dd$ phase 
\cite{Fujimoto:2019sxg}; 
in addition to the conventional 2SC phase, a $P$-wave 
condensation  $\left< d\nabla d \right>$ is suggested.
Vortex structures in the 2SC+$dd$ phase were studied in 
Refs.~\cite{Fujimoto:2020dsa,Fujimoto:2021bes,Fujimoto:2021wsr} 
in which an exotic vortex called a non-Abelian Alice string was found. 
In particular, in Ref.~\cite{Fujimoto:2021bes}, 
how vortices in nuclear and quark matter should be 
connected along the crossover.
It will be interesting whether core structures found in 
this paper are preserved or deformed 
along this connection.

Finally, a novel type of the Berezinskii-Kosterlitz-Thouless (BKT) transition 
of vortex molecules in two-component systems 
was reported
in Ref.~\cite{Kobayashi:2018ezm}, 
and thus the BKT transition should be investigated in 
neutron $^3P_2$ superfluids.

\section*{Acknowledgment}

We would like to thank Yusuke Masaki for helpful discussions and comments.
The work of M.K. is partly supported by JSPS KAKENHI (Grants No. 20K03765, No. 19KK0066), and by Osaka City University Advanced Mathematical Institute (MEXT Joint Usage/Research Center on
Mathematics and Theoretical Physics JPMXP0619217849).
The work of M.N. is supported in part by JSPS KAKENHI (Grant No.~JP18H01217).

\newpage

\bibliography{vortex}

\end{document}